# SPACE-EFFICIENT K-MER ALGORITHM FOR GENERALISED SUFFIX TREE

Freeson Kaniwa, Venu Madhav Kuthadi, Otlhapile Dinakenyane
and Heiko Schroeder

Botswana International University of Science and Technology, Botswana.

### ABSTRACT

*Suffix trees have emerged to be very fast for pattern searching yielding O (m) time, where m is the pattern size. Unfortunately their high memory requirements make it impractical to work with huge amounts of data. We present a memory efficient algorithm of a generalized suffix tree which reduces the space size by a factor of 10 when the size of the pattern is known beforehand. Experiments on the chromosomes and Pizza&Chili corpus show significant advantages of our algorithm over standard linear time suffix tree construction in terms of memory usage for pattern searching.*

## 1. INTRODUCTION

Suffix trees are well known and powerful data structures that have many applications [1-3]. A suffix tree has various applications such as string matching, longest common substring, substring matching, index-at, compression and genome analysis. In recent years it has being applied to bioinformatics in genome analysis which involves analyzing huge data sequences. This includes, finding a short sample sequence in another sequence, comparing two sequences or finding some types of repeats. In biology, repeats found in genomic sequences are broadly classified into two, thus interspersed (spaced to each other) and tandem repeats (occur next to each other). Tandem repeats are further classified as micro-satellites, mini-satellites and satellites. Micro-satellites are of size 2-5 characters, mini-satellites are of sizes 5-100 characters and satellites are of anything greater than 100 [4]. Each of these repeats have been found to have their biological functions. Suffix trees have been studied many times under different names and that possibly makes them more important and useful for instance, the prefix tree [5], the PAT tree [6], repetition finder [7], position tree [8-10], sub-word tree [5, 11] and compacted bi-tree [1].The most attractive property of the suffix tree is that, after construction, its able to solve most problems in O (*m*) time (where *m* represents substring size) [12].

Due to the usefulness of suffix trees numerous sequential algorithms have been proposed [2, 13-18]. However, all these algorithms have been shown to suffer from high memory requirements (approx. 20 to 50 times the size of the string to be indexed) [19]. This paper therefore addresses the problem of high memory requirements in suffix trees.

This paper proposes a variant of generalized suffix tree. This data structure is a trie of all the *k*-mers, this has been proposed before however ours differ from these since it does not have suffices of the *k*-mers [20,21]. This data structure is truncated to depth *k* to store *k*-mers. Therefore the proposed algorithm makes generalized suffix tree of all k-length substrings of input string S with only *k*-length substrings and with a space savings of a factor of 10.

The remaining part of the paper is structured as follows; we introduce the Preliminaries, that is, the background and notations of the concepts used in the entire paper in Section II. We provide brief literature review of related work in Section III and present our space-efficient *k*-mer search algorithm in section IV and then consequently discuss the analysis of our results in Section V and VI. We provide conclusion in Section VII.

## 2. PRELIMINARIES

Let $\Sigma$ be a finite ordered *alphabet* on a string $S$. A string $S = s_1 s_2 s_3 \ldots s_n$ are characters in a sequence from an alphabet $\Sigma$ and $\$ \notin \Sigma$, $\$ $ is the end-of-string symbol. A substring of $S$ is a string $S_{i..j} = s_i s_{i+1} \ldots s_j$, where $1 \leq i \leq j \leq n$. A pattern is a substring over the alphabet $\Sigma$. A pattern $P = p_1 p_2 p_3 \ldots p_k$ occurs at position $j$ of string $S$ iff $p_1 = s_j$, $p_2 = s_{j+1}, \ldots p_k = s_{j+k-1}$. A prefix of $S$ is any short string $S_{1..j}$ where $1 \leq j \leq n$. A suffix of $S$ is any substring $S_{i..n}$, where $1 \leq i \leq n$ [22]. We use $k$ to represent the size of the pattern to be searched hence $k$ - mer. In this paper, we define a pattern as a substring to be searched in a given string.

Generally a suffix tree for a string $\delta$ is a rooted tree with edges labeled with non-empty substrings of $\delta$. Vertices represents suffixes and the root is represented by an empty string. Every substring $\alpha$ can be represented as location pair $(\beta, \gamma)$ where $\gamma$ is the suffix of string $\alpha$ and $\beta$ is the node of the suffix representing a prefix of string $\alpha$. To make the shape of the suffix tree, the following two rules should be maintained:

1. No parent has two edges to its children with labels beginning the same character.
2. There is no non-root node in the tree with only one child.

For example Fig. 1 shows construction transition of a suffix tree from *oo* to *ook*.

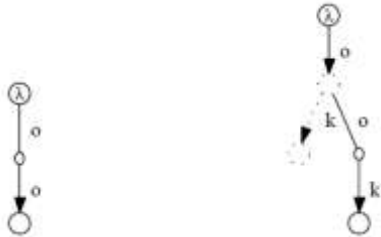

Fig. 1: A suffix tree construction transition example [23]

It is a standard practice to append the string with $ as the terminal symbol, so as to generate an explicit suffix tree, as shown in Fig. 2. Fig. 2 shows a complete suffix tree for the string $\delta$ = *xabxac* with its suffixes as {*xabxac*$, *abxac*$, *bxac*$, *xac*$, *ac*$, *c*$, $}

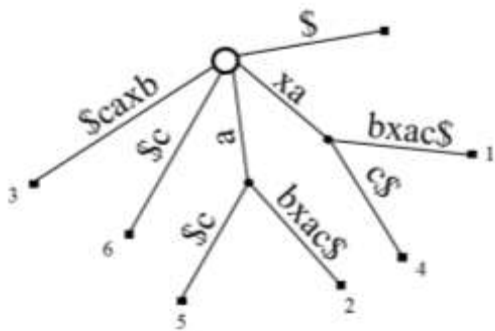

Figure 2: Explicit Suffix Tree of δ

 A generalized suffix tree is a type of suffix tree constructed by indexing multiple strings {S₁,...,Sₙ} for *n* strings, whereas a standard suffix tree is constructed by indexing one string. In this paper we use a generalized suffix tree.

## 3. RELATED WORK

We briefly review the existing algorithms for construction of suffix trees. A lot of work relating to optimized construction of suffix tree algorithms has been done [2, 13-18]. There are two broad categories of suffix tree construction algorithms, which are, In-memory and Disk-based. In-memory is normally applied when both the string and the suffix tree can fit in the internal memory. Disk-based algorithms work by extending memory to the secondary storage, that is, the hard disk or any external memory by storing some parts of the string *S* or suffix tree on disk memory which cannot fit in the main memory. However all these algorithms tries to address one major drawback of suffix trees, that is of high memory requirements, which makes it impractical to be applied in huge data sequences [1, 19, 24-29]. This calls for new approaches needed to deal with this issue of high memory requirements in suffix trees [28].

The initial algorithm for suffix tree construction was done by Weiner [1].The algorithm runs in O (*m* log ∑) time. McCreight [2] improved this algorithm by providing a more efficient construction within the same asymptotic bound. Manber and Myers [30] and independently Gonnet et al. [6] introduced suffix arrays as a space − efficient alternative data structure but still related to suffix trees, thus trying to address the drawback of high memory requirement. However suffix trees still have superior features over suffix arrays such as inexact searches, inference of motifs etc. these operations are still possible on the suffix arrays but are burdensome on suffix arrays compared to suffix trees [31, 33]

Stoye and Gusfield [33], proposed an algorithm which runs in O (*n* log *n*) time and requires O (*n*) space. The algorithm is used to retrieve all occurrences of tandem repeats. Abouelhoda [25], improved on the Stoye and Gusfield algorithm limitations by introducing a new data structure called the suffix array which did improve the space requirements however this suffix array data structure does not cope well when for retrieving approximate repeats and inference of motifs as stated earlier which makes it impossible to apply it in situations which needs such operations.

Valimaki et. al. [22] designed an algorithm to address the memory challenge by using the Compressed Suffix Tree (CST), there was an improvement on the memory requirements however the memory requirements remained impractically high for it to be adopted and usable. For instance, for the whole human genome, the indexing takes around 4 days, the complete tree takes around 8.5 GB with a maximum amount of memory consumption of 24 GB. It is also important to know that this construction time was achieved on a 32 GB-memory computer which completely shows the challenge of high memory requirements.

# 4. SUFFIX TREE CONSTRUCTION

We now describe our algorithm for suffix tree construction with the help of an example, shown in Example 1. Our algorithm constructs the suffix tree by looking at the pattern size ($k$) and then employs the sliding window technique (as shown in Fig. 5) to construct the tree with a shift of one character at a time using a window size of $k$. We make use of the Example 1 to explain our idea.

Example 1: Given a string $S = \{ACGTCCTGG\}$ and $k = 4$ (where $k$ is the pattern size), the algorithm first extract the size $k$ substring from $S\$$ based on the sliding window approach as shown in Fig. 5. Each window is a substring, hence set of substrings $sub = \{ACGT\$, CGTC\$, GTCC\$, TCCT\$, CCTG\$, CTGG\$\}$. Then, a generalized Suffix Tree is constructed from the set $sub$. The final tree has a maximum height of 4 since $k$ is equal to 4. Each edge starting from the root to leaf is of length 4. The complete suffix tree for $S$ is shown in Fig. 3.

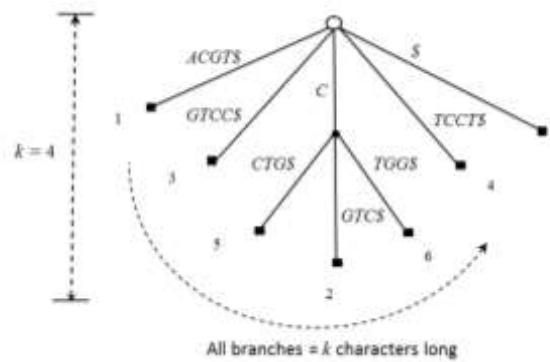

Figure 3: 4-mer Suffix Tree of S

Every pattern with $k = 4$ can be retrieved in O ($|p|$) time, where $p$ is the pattern to be searched. The size of the tree has reduced to height $k$ which reduces space requirements as compared to the complete suffix tree shown in Fig. 4.

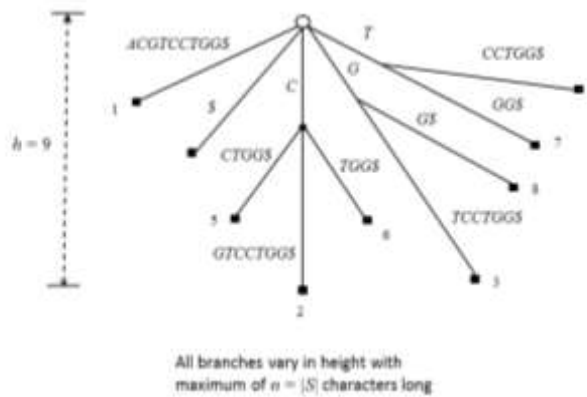

Figure 4: Complete Suffix Tree

Finally the algorithm creates the suffix tree $k$ - substring by $k$ - substring to come up with the complete suffix tree for the string $S$.

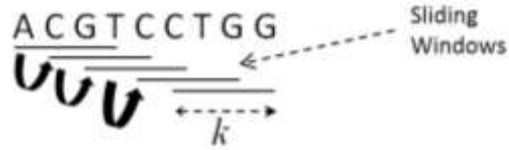

Figure 5: Sliding windows technique illustration

Fig. 6 shows a high-level view of our idea, as shown in the diagram, the standard suffix tree will always be larger than our generalized suffix tree, since our suffix tree goes up to a height of $k$, where $k$ is the size of the repeat to be searched. The standard suffix tree goes up to varying heights, but to a maximum of $n$, where $n$ is the size of the complete string to be indexed. In general, our tree represents a chopped suffix tree with the same capability to retrieve all repeats of size $k$ (Figure 6, Part b).

We now describe our algorithm (Algorithm 1) which we call KSA (K-*mer* Search Algorithm). Our high level algorithm starts by sliding a window of size $k$ and generating a substring of $k$ fixed length (*sub*), this executes up to the last index, that is, $n - k$ index (lines 5 - 6). The next stage in the same for loop involves passing the just generated $k$ - substring to the *insert* subroutine for adding it to the generalized suffix tree and this runs up to $n - k$ steps (line 7). $k$ - substring is then inserted into the tree *ST*, node by node (lines 8 - 10). The insert function works by inserting the substring as in any generalized suffix tree however the suffixes of the substrings are not included. These substrings are generated by a sliding window method. At the end, a complete generalized suffix tree is constructed of $k$ - height and not $n$ - height which gives our suffix tree huge $n - k$ memory savings.

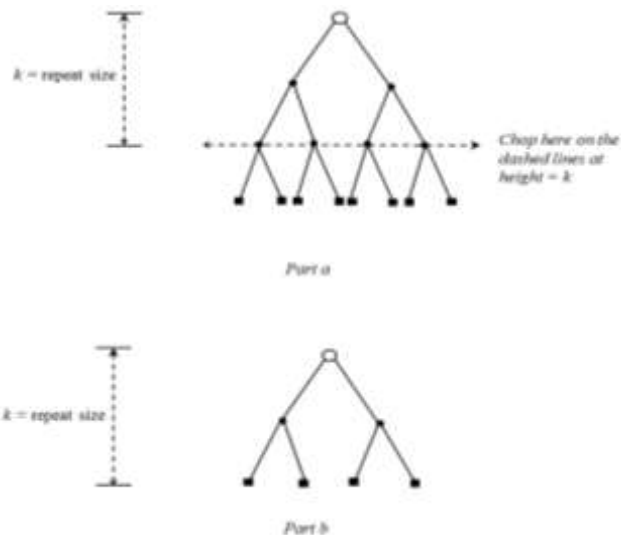

Figure 6: Part a) Standard suffix tree vs. Part b) our generalized suffix tree

**Algorithm 1**. Algorithm K-mer Search Algorithm (KSA)
**Input**   : S,$n$,K
**Output** : A suffix tree representation ST of S of height $k$

    1 ST :={ }

    2 $n$ :=|S|

3 k := K  //size of the repeat

3 ST :={ }

4 *sub* := [ ]

5 for $i \leftarrow 0$ to $n - k$ do

6　　　*sub* = S [*i, i + k* - 1]  //sliding *k*-size window

7　　　ST.insert (*sub*)

8 insert:

9　　　for $i \leftarrow 0$ to |*sub*| - 1do

10　　　　　add.node [*sub* [*i*] ]

11 return ST

## 5. COMPLEXITY ANALYSIS

Theorem 5.1. KSA takes O (*n*) time and O (*n*) space where *n* is the size of the string and *k* is the length of the pattern.

Proof: Our algorithm executes $n - k$ times in the *for loop* (lines 5 - 8). Then executes |*sub*| - 1 times, which is *k* times since the size of the substring is *k*. Then we have ((*n* -*k*) *k*) times (*k* times for the insert function since it is in the inner loop which finally gives us O (*nk*) and hence O (*n*) since *k* is a constant.. If each symbol uses one unit of memory and each branch is *k* long that means, with $n - k$ substrings. Then the space complexity is O (*n* ) since *k* is a constant.

## 6. IMPLEMENTATION AND EXPERIMENTS

We implemented the algorithm in Python. We implemented every part of the algorithm. We use pythons' version of Mark Carlson C++ code for comparative analysis which is based on the Ukkonen linear time algorithm which is suffix tree based (ST-based) [34, 35]. Our implementation is complete since it supports pattern searching in string of any size as big as the human genome and therefore is ready for use.

### 6.1 Experimental environment

Development was done using Python 3.5.2 and Anaconda 4.1.1 to implement the algorithm in our experiments. The experiments were conducted on a HP Z820 Workstation with a 3.30 GHz sixteen-core Intel Xeon E5-2623 chip which has 10 MB L3 Cache, and we did not apply any parallellism. The workstation is installed with Ubuntu 14.04LTS (64-bit) operating system with 32 GB main memory, swap space of 64 GB, and 1 TB Serial ATA Hard Disk Drive at 7,200 RPM. We used genomic data (real-world) to test the space usage and usability of our KSA algorithm. We got rid of all the *N*s characters (masked N), so the sequences only contain original characters of the sequence.

1. The human *X*-Chromosome sequences sourced from NCBI [36].
2. The complete human genome sequence [36].
3. Protein sequences sourced from the Pizza&Chili Corpus. [37].
4. Rice Chromosome 12 (Genbank accession AP008219), we chose the rice chromosome 12 sequence because chromosome 12 is the most repeat rich chromosome for this plant [38].

Table 1. Space usage comparison between ST-based algorithm [34] and our algorithm

| Dataset | Text Size (MB) | ST-Based (KB) | Our Algorithm (KB) |
|---------|----------------|---------------|--------------------|
| Rice 1  | 5  | 7,523,392   | 716,224   |
| Rice 2  | 10 | 15,064,532  | 935,572   |
| Rice 3  | 15 | 23,080,864  | 1,103,240 |
| Rice 4  | 20 | >32,000,000 | 1,219,996 |
| Rice 5  | 25 | >32,000,000 | 1,311,764 |
| Prot. 1 | 10 | 15,397,672  | 706,852   |
| Prot. 2 | 20 | >32,000,000 | 898,200   |
| Prot. 3 | 30 | >32,000,000 | 1,058,024 |
| Prot. 4 | 40 | >32,000,000 | 1,134,716 |
| Prot. 5 | 50 | >32,000,000 | 1,180,444 |

We used megabytes to measure the Text size. A 15MB text size file contains approximately 15 million characters. Data are stored using 1 byte per character. The ones *'> 32,000,000'* are estimates since our main memory is 32GB and the program did not terminate possibly due to insufficient main memory. The construction for our algorithm is for 10-*mer*, that is, for size ten patterns.

Table 2. Space usage of for complete chromosomes and the whole genome

| Dataset | Text Size (MB) | ST-Based (bytes) | Our Algorithm (KB) | | |
|---------|----------------|------------------|----------|----------|-----------|
| | | | $k = 5$ | $k = 10$ | $k = 15$ |
| Rice Chr. 12 | ≈ 21.5     | >32,000,000 | 43,780    | 1,257,628   | 23,281,280  |
| X-Chr.       | ≈ 151,855  | >32,000,000 | 159,488   | 1,429,320   | >32,000,000 |
| E.H.G.       | ≈ 2,992,718| >32,000,000 | 5,851,588 | >32,000,000 | >32,000,000 |

*Rice Chr. 12* represents Rice Chromosome 12, *X-Chr.* represents the Human Chromosome X and *E.H.G.* represents the Entire Human Genome. All columns show memory usage in bytes except for the Text Size column. We show the peak memory usage for our construction basing on pattern size. We used *memusg* tool to measure the peak memory usage which is a 'time'-like utility for Unix that measures peak memory usage [39], and to calculate the programs' execution time in seconds we used system time, shown in Table 3.

Table 3. Time cost for complete chromosomes and the whole genome

| Dataset | Text Size (MB) | ST-Based (seconds) | Our Algorithm (seconds) | | |
|---------|----------------|--------------------|----------|----------|-----------|
| | | | $k = 5$ | $k = 10$ | $k = 15$ |
| Rice Chr. 12 | ≈ 21.5      | n/a | 40.263     | 91.090  | 274.414 |
| X-Chr.       | ≈ 151,855   | n/a | 279.484    | 584.902 | n/a     |
| E.H.G.       | ≈ 2,992,718 | n/a | 55,380.870 | n/a     | n/a     |

## 6.2 Main observation

We compared the memory usage of our algorithm with the memory usage of the Suffix Tree-based method [34]. The following main observations were noted from this experimental study:

1. The ST-based algorithm uses more than 10 times the size of our KSA algorithm memory usage. This was shown on the Rice and Protein data (Table. 1). These memory savings makes it possible to construct the generalized suffix tree for the entire human genome with just less than 6 GB. Our algorithm can therefore fit in a standard computer with 6 GB main memory to find the requested pattern of size $0 - 10$ characters long.

2. When the size of the pattern exceeds 15 on the *X*-Chromosome and the entire human genome, our algorithm requires more internal memory, that is, more than 32GB which was used in this experimental study.

3. We also noted that, the time aspect of the construction of all the datasets was less than 5 minutes except for the entire human genome which took approximately 15 hours to complete construction and to search for the pattern (Table. 3).

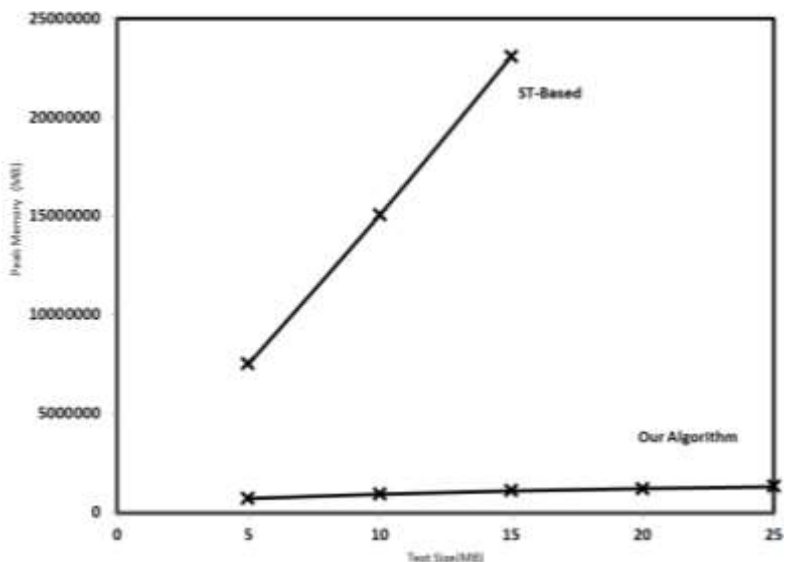

Figure 7: Space usage comparison on Rice sequences

Fig. 7 shows memory usage between the ST-based method and our KSA algorithm. Memory is measured in bytes. The ST-based method requires peak memory of 24GB for the 15Mb of Rice test data (Table. 1) whereas ours consume only 1.2GB for the same dataset size. Fig. 8 shows peak memory usage between the ST-based algorithm and our algorithm. Memory is measured in bytes; one byte per base is used to store data. The ST-based method could not terminate for anything greater than 20MB text size (with $k = 10$). This is possibly because it required more main memory than the available 32GB as shown from the memory usage of 10MB text size. Our algorithm used only under 2GB for the 20MB which could complete with the ST-based method.

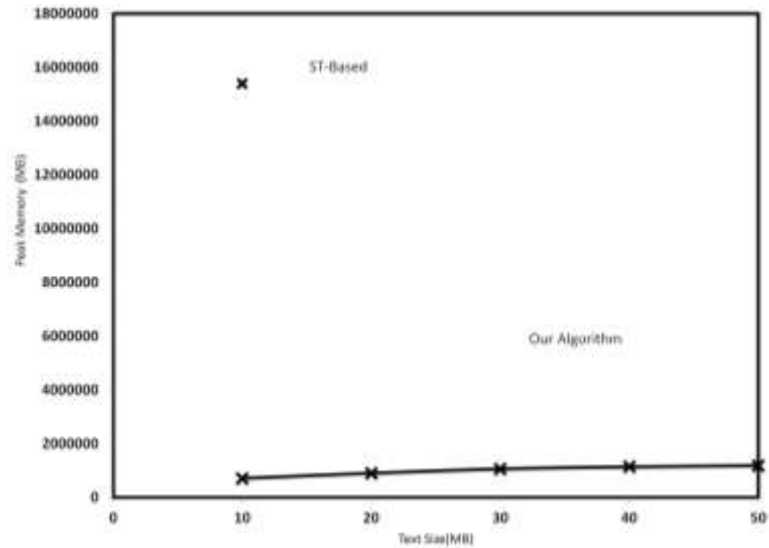

Figure 8: Memory usage on Protein sequences

# 7. CONCLUSION

In this paper we proposed a data structure which is a variant of a generalized suffix tree and have that the biggest drawback of the suffix tree is of high memory requirements. Through our experiments we have managed to reduce the memory requirements by a factor of the size of the repeat to be searched. We have also provided a space efficient solution to suffix tree, that is, if the pattern to be searched is short (0 - 10) enabling us to index the entire human genome with only 6GB of main memory. Our approach would be useful in instances where the repeats to be searched are small such as the micro-satellites where both the sequence and the tree can both fit in the main memory. Therefore, future work would involve considering disk-based approaches to augment the part which does not fit in the internal memory and consideration of parallelization to improve running time.

# 8. ACKNOWLEDGMENTS

We would like to thank the IT Department of BIUST for providing the HPZ820 workstation for the test runs.